\documentstyle[12pt,aaspp4]{article}
\begin{document}

\title{The Inner Boundary Condition for a Thin Disk Accreting
Into a Black Hole}

\author{B. Paczy\'nski}
\affil{Princeton University Observatory, Princeton, NJ 08544--1001, USA}
\affil{E-mail: bp@astro.princeton.edu}

\begin{abstract}
Contrary to some recent claims the `no torque inner boundary condition'
as applied at the marginally stable orbit is correct for geometrically
thin disks accreting into black holes.
\end{abstract}


A transition from a thin accretion disk to a stream freely falling into a
black hole was a topic of many papers about two decades ago.  That ancient
work is very well reviewed in the introduction of Abramowicz \& Kato (1989).
It was well established that there is a transition from a subsonic radial
accretion flow in the nearly Keplerian disk to a transonic flow near the
marginally stable orbit at the radius $ r_{in} \approx r_{ms} $, and a free
fall into the central black hole for $ r < r_{in} $.  The free fall proceeds
with a conservation of angular momentum, hence the streamlines are spiral.
It was shown that the `no torque inner boundary condition'
is an excellent approximation at $ r_{in} $.  The reason was simple:
no information could propagate upstream in the supersonic region inwards
of $ r_{in} $.

Recently, the `no torque inner boundary condition' has been questioned
(Krolik 1999, Gammie 1999, Agol \& Krolik 2000, to be referred to as KGA).
The authors have chosen to ignore Abramowicz \& Kato (1989), and the
many references therein.  They claim that magnetic torques provide
a strong interaction between the stream and the disk, and generate a strong
torque at $ r_{in} $.  This is a puzzling result, as the flow is
transonic near $ r_{in} $ according to KGA, and the radial infall
becomes supersonic inwards of $ r_{in} $.

The subsonic accretion for $ r > r_{in} $ is not discussed by KGA, but their
disks seem to be described by more or less standard `alpha models', with
the $ \alpha $ parameter due to tangled magnetic fields, following the work 
of Balbus \& Hawley (1998).  While the differential rotation winds up the
magnetic field lines and perhaps gives rise to a dynamo, the magnetic
energy density is kept in equilibrium, at the level roughly $ \alpha $ times
the gas and/or radiation energy density.  This implies that the reconnections
and/or the escape of magnetic flux to a corona prevent the magnetic
field from growing up to an equipartition with the gravitational binding
energy, and prevent the disk from becoming geometrically thick.  In other 
words, the assumption that the disk is geometrically thin for $ r > r_{in} $
implies that magnetic energy is efficiently dissipated there.

Whatever are the physical processes responsible for the dissipation of magnetic
energy in the disk, KGA assumed that they do not operate in the stream,
inwards of $ r_{in} $.  Hence, the differential rotation builds up magnetic 
energy in the matter falling into the black hole up to the level comparable 
with the gravitational binding energy, the Alfv\'en velocity becomes 
relativistic and the information can travel upstream, all the way to $ r_{in} $.

What is wrong with this picture?  First of all it is not
clear at all why the physics of the magnetic field -- plasma interaction
should be so dramatically different in the disk ($ r > r_{in} $) and in
the stream ($ r < r_{in} $)?  From the local point of view of a blob of gas
with tangled magnetic field, there is hardly a difference between the flow
on the two sides of the sonic point, with the matter accreting along a tight
spiral.  The character of the flow changes dramatically only when the
radial infall velocity becomes comparable to the rotational velocity.

Let us make a quantitative analysis of the flow pattern using a pseudo --
Newtonian approximation (Paczy\'nski \& Wiita, 1980).  Let us define
the parameter $ \alpha $ with a more or less standard formula for the
kinematic viscosity $ \nu $:
$$
\nu = \alpha c_s H = \alpha H^2 \Omega ,
\eqno(1)
$$
(see for example eq. 2.9 of Chen, Abramowicz \& Lasota, 1997 $ \equiv $ CAL),
where $ H $ is the disk thickness, $ \Omega $ is angular velocity, and $ c_s $
is the effective sound speed; if the magnetic pressure dominates then $ c_s $
is the Alfv\'en speed.  The equation of angular momentum balance for a disk in
a steady -- state accretion can be written as
$$
v_r = \alpha ~ H^2 ~ { l \over l - l_0 } ~ { d \Omega \over dr }
\approx \alpha ~ H^2 ~ { l \over l - l_0 } ~ { \Omega \over r }
\approx \alpha ~ v_{rot} \left( { H \over r } \right) ^2 ~ { l \over l - l_0 } ,
\eqno(2)
$$
(see eq. 2.4 of CAL),
where $ v_r $ is the radial velocity, $ v_{rot} $ is the rotational
velocity, $ l(r) $ is the specific angular momentum
at radius $ r $, and $ l_0 $ is the integration constant, corresponding to
the asymptotic angular momentum at the inner end of the flow.  The equation (2)
does not assume that the radial velocity is small, i.e. it holds within the
disk as well as within the stream.  Far out in the disk, where $ l \gg l_0 $,
we obtain the well known formula
$$
v_r \approx \alpha ~ v_{rot} \left( { H \over r } \right) ^2 ,
\hskip 1.0cm r \gg r_{in} .
\eqno(3)
$$

At the sonic point we have $ v_r = v_s \approx v_{rot} ~ H/r $, and the
equation (2) becomes:
$$
{ v_r \over v_s} = 1 \approx \alpha ~ { H_{in} \over r_{in} } ~ 
{ l_{in} \over l_{in} - l_0 } ,
\hskip 1.0cm r = r_{in} .
\eqno(4)
$$
If the disk is thin, i.e. $ H_{in} / r_{in} \ll 1 $, and the viscosity is 
small, i.e.  $ \alpha \ll 1 $, as assumed by Gammie (1999), then the eq. (4)
implies that $ (l_{in} - l_0 ) / l_{in} \ll 1 $, i.e. the specific angular
momentum at the sonic point is almost equal to the asymptotic angular
momentum.  

In a steady state disk the torque $ g $ has to satisfy the equation of
angular momentum conservation, which can be written as
$$
g = \dot M \left( l - l_0 \right) , \hskip 1.0cm 
g_{in} = \dot M \left( l_{in} - l_0 \right) .
\eqno(5)
$$
It is clear that for a thin, low viscosity disk the `no torque inner boundary
condition' is an excellent approximation, as established two decades ago
(see Abramowicz \& Kato, 1989).  However, if the disk and the stream are
thick, i.e.  $ H/r \sim 1 $, and the viscosity is high, i.e. 
$ \alpha \sim 1 $, then the angular momentum varies also in the stream,
as demonstrated by Chen, Abramowicz \& Lasota (1997), in accordance 
with the simple reasoning presented above.

Note, that it does not matter how complicated is the stream,
the conditions (4) and (5) at the sonic point cannot be affected by what
happens in the supersonic flow, even if the $ \alpha $
parameter changes a lot between $ r_{in} $ and $ r \ll r_{in} $.
The KGA claim that the `no torque inner boundary condition' is contradicted
by their models implies that their models do not satisfy the angular momentum 
conservation law, on which the reasoning of this paper is based.

I am very grateful to Dr. M. Abramowicz and Dr. J. P. Lasota for many 
important comments.  This work was not supported by any grant.


\end{document}